\documentclass[aps,pre,preprint,amsmath,amsfonts,showpacs]{revtex4-1}
\usepackage[utf8]{inputenc}
\usepackage[T1]{fontenc}
\usepackage[english]{babel}
\usepackage{amssymb}
\usepackage{graphicx}
\usepackage{multirow}
\usepackage{endnotes}
\let\footnote=\endnote

\begin{document}
\title[KMC Simulations of Proton Conductivity]{Kinetic Monte Carlo Simulations of Proton Conductivity}
\author{T.~Masłowski$^{1a}$\footnotetext{T.~Maslowski@if.uz.zgora.pl}, A.~Drzewiński$^1$, J.~Ulner$^2$, 
J.~Wojtkiewicz$^3$, M.~Zdanowska-Frączek$^4$, K.~Nordlund$^{5}$ and A.~Kuronen$^{5}$}
\affiliation{$^{1}$Institute of Physics, University of Zielona Góra, ul. Prof. Szafrana 4a, 65-516 Zielona~Góra, Poland}
\affiliation{$^{2}$Institute of Low Temperature and Structure Research PAN, ul. Okólna 2, 50-422 Wrocław, Poland}
\affiliation{$^{3}$Department of Mathematical Methods in Physics, Faculty of Physics, University of Warsaw,
ul. Hoża 74, 00-682 Warszawa, Poland}
\affiliation{$^{4}$Institute of Molecular Physics, Polish Academy of Sciences, ul. M. Smoluchowskiego 17, 60-179 Poznań, Poland}
\affiliation{$^{5}$Division of Materials Physics, Department of Physics, P.O. Box 43, FI-00014, University of
Helsinki, Finland}

\date{\today}

\begin{abstract}
The kinetic Monte Carlo method is used to model the dynamic properties of proton diffusion in anhydrous proton
conductors. The results have been discussed with reference to a two-step process called the Grotthuss
mechanism. There is a widespread belief that this mechanism is responsible for fast proton mobility. We showed in
detail that the relative frequency of reorientation and diffusion processes is crucial for the conductivity.
Moreover, the current dependence on proton concentration has been analyzed. In order to test our microscopic model
the proton transport in polymer electrolyte membranes based on benzimidazole C$_7$H$_6$N$_2$ molecules is studied.  

\end{abstract}

\maketitle

\section{Introduction}

Proton transfer is of great general importance to many processes in chemical and bio-chemical reactions.
Historically,
it appeared first in the context of the fast proton charge transport in water and ice. What is crucial is that the high
mobility of the proton stems from the fact that it does not move freely but is passed by successive water molecules
via the so-called Grotthuss mechanism \cite{Grotthuss}.

Recently the polymeric systems which conduct protons in the absence of any water have become the subject of intensive research. 
This can be associated with the fact that proton conductivity of some water containing compounds suffers from
substantial proton conductivity decrease with decreasing degree of hydration. In most cases it takes place at
temperatures close to the boiling point of water ($373.15$ K). So, the promising strategy is to substitute water with
a high boiling proton solvent (e.g. the benzimidazole with the melting temperature $447$ K). There are also other anhydrous
proton conductors as the solid acids with the formula $MH_nXO_4$, where $M$ is a metal like Cs, K, Rb or an organic
monovalent cation and $XO_4$ is the tetrahedral anionic group, where $X=$S,Se,P,As \cite{Kreuer,Munch}. In the phase
with high conductivity they exhibit anhydrous proton transport with conductivities of the order of $10^{-2}$ Scm$^{-1}$
at the temperature of about 400--450 K.

There have been many attempts to describe the properties of proton conductors using the soliton approach 
\cite{Yomosa,Pang,Gordon}, the polaron mechanism \cite{Pavlenko,Stasiuk}, the MD
calculation \cite{Chisholm,Wood}, and recently the kinetic Monte Carlo (KMC) method \cite{Hermet}. Although the description of the mechanism
of proton mobility still cannot be regarded as satisfactory, it seems that the key elements are common for a wide
range of compounds. In a similar fashion to the proton conductivity in water they are realized by a two-stage
mechanism \cite{Kreuer,Pavlenko} consisting of thermally induced structural reorganization (e.g., rotations of 
the tetrahedra for the solid acids) and proton tunneling in hydrogen bonds (H-bonds).

Because the diffusion of protons is performed along hydrogen-bond networks whose dimensionality varies from 0 to 3
\cite{zerod,struct,Chis2,Chis3,Nelm}, a low-dimensional model can also be a good candidate for realistic compounds 
\cite{Chis2}. An example is the microscopic model introduced by Pavlenko and Stasyuk \cite{Pavlenko,Stasiuk} where 
besides the proton transport mechanism, the effect of displacement of the nearest oxygens during hydrogen-bond 
formation is also introduced, leading to the polaronic effect. In this quantum mechanical model a two-stage mechanism 
is realized in a zigzag hydrogen-bonded chain by the creation and annihilation of quasiparticles with two transfer
parameters corresponding to rotations and tunnelings. Unfortunately, computational difficulties require additional
simplifications, such as the use of linear response Kubo theory, but even then only small systems can be examined.

Then a natural way to explore the Grotthuss mechanism is to use numerical simulations that have become an indispensable
tool for the investigation of various physical processes. One of the principal methods is molecular dynamics simulations
which are very often applied to mass transport problems, with time scale of the order of nanoseconds. However, to 
achieve the typical
time scale for proton transport \cite{Hermet} the time scale of the order of microseconds is required. Such time
scales are not accessible to conventional molecular dynamics, but can be accessed with the KMC
approach \cite{KMC,You66,Fic91,Nur00,Dju09a}. Moreover, the KMC-based simulations are simple enough to effectively test
the hypothesis arising from the experiment but they are also capable of covering all the necessary constituents
responsible for protons dynamics. 

The main aim of our paper is to propose the microscopic model of proton conductivity in anhydrous proton conductors,
such as polymeric systems or solid acids. In order to verify its usefulness, proton conductivity results have been
compared with the experimental data for a polycrystalline sample of the benzimidazole. Our research can shed some light
on proton mobility in anhydrous systems.

\section{The model}

Since the proton diffusion process may be divided into sub-processes separated in time and localized in space, as is
the case of the Grotthuss mechanism, the KMC method is a natural choice for the analysis of phenomena during
protons flow. As the model system we propose a chain of parallel rigid rods whose ends can be occupied by protons,
one proton per end. Rods with or without protons can independently rotate by the angle $\pi$.

\begin{figure}[htb]
\centering
\includegraphics[width=8.5cm]{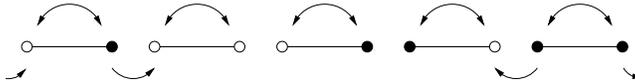}
\caption{A possible distribution of protons (solid dots) and allowed movements in the chain.
Upper arrows represent rod rotations by angle $\pi$ that may but do not have to lead 
to a different configuration (this happens if the rotating rod is not occupied or occupied 
by two protons). Lower arrows represent acceptable hoppings of a proton between the neighboring rods for
the particular configuration in the picture. The periodic boundary conditions permit a hop from
the rightmost rod to the leftmost one.}
\label{plot01}
\end{figure}

Protons can also migrate by hopping from one rod to the nearest one provided the end of the adjacent rod is empty 
(see Fig.~\ref{plot01}). Rods should be considered as, e.g., benzimidazole molecules making the $180^\circ$ flip or
the one-dimensional realization of tetrahedral anionic groups in the solid acids. In turn, the hopping from one rod
to the neighboring one corresponds to the transfer of a proton in a hydrogen bond which is created between 
electronegative atoms of neighboring anionic groups.

The number of protons in the system may be freely adjusted from 0 to $2N$, where $N$ is the number of rods. It
gives us more flexibility than is possible in nature where only specific concentrations of protons are realized
\cite{struct,Chis2,Chis3,Nelm}. By the proton concentration we mean the ratio $c=n/(2N)$, where $n$ is the number
of protons.

In the presence of the external electric field the proton diffusion is ordered. To make the current flow possible 
the periodic boundary conditions are imposed. The KMC method yields time evolution of the system, thus if we count 
protons crossing a specified position in a chain then we are able to calculate the proton current. 
At this stage of our considerations only dc current is considered.

\subsection{Kinetic Monte Carlo}

The time-evolution of the system is realized by a jump of a particle from one local energy minimum to another.
For this purpose one needs to know {\it a priori} all transition rates from every configuration to every other allowed
one \cite{Fic91}. It may happen that after a transition the system will be in the same configuration, e.g., when
a rod without protons rotates.

\begin{figure}[htb]
\centering
\includegraphics[width=8.4cm,height=5.19cm]{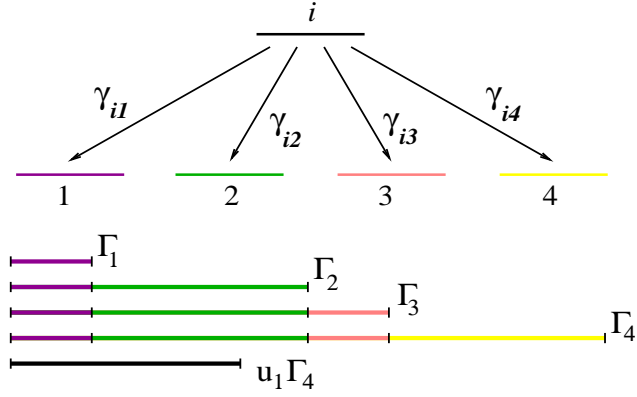}
\caption{(Color online) A schematic description of the new configuration choice in the KMC algorithm.
For the situation presented in the picture, provided that a number $u_1$ will be drawn 
the system is transformed to configuration number 2.}
\label{plot02}
\end{figure}

When all allowed configurations and all transition rates are known the KMC method gives the answer to the questions 
of how long the system remains in the same configuration and to what configuration it will evolve \cite{KMC}.
If we denote by $\gamma_{ij}$ the transition rate from configuration $i$ to $j$ and define $\Gamma^i_n=\sum_j^n \gamma_{ij}$
then the system will be transformed to configuration $l$ satisfying the following relation
\begin{equation}
\Gamma^i_{l-1} < u_1\Gamma^i_{N_i} \le \Gamma^i_l\;, \label{orderN}
\end{equation}
where $N_i$ is the number of all possible configurations accessible from $i$ and $0<u_1 \leq 1$ is a number from
the uniform distribution that has to be generated (see Fig.~\ref{plot02}). 

The selection of a new configuration using Eq. (\ref{orderN}) costs the time of order
$O(N_i/2)$, but we may speed up this process significantly by applying the binning method \cite{binning,Blue} for
the KMC algorithm. In this case transition rates are stored on the special binary tree which reduces the computational 
time to the order of $\log_2N_i$.

Another uniform random number, $u_2$, is necessary to determine the life-time of the configuration $i$ using
the following formula:
\begin{equation}
\Delta t = - \frac{\log u_2}{\Gamma_{N_i}}\;,
\end{equation}
according to the assumption that the lifetime follows the Poisson distribution, which is a manifestation of
the presumption that all transitions are independent. When the new configuration $l$ is chosen we repeat the above
steps treating $l$ as the starting configuration.

\subsection{Bjerrum $D$ and $L$ defects}

\begin{figure}[htb]
\centering
\includegraphics[width=8.4cm]{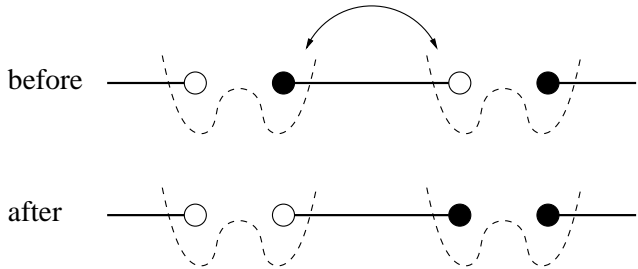}
\caption{There are only two configurations requiring the additional factor representing effective
Coulomb forces: one presented in the picture above and its mirror reflection. The dashed lines represent
H-bond potentials. For the initial configuration above with one proton in each H-bond before the rod's rotation 
there is one proton and one vacancy in each H-bond which is energetically favorable.
After the rotation two protons meet in one H-bond and two vacancies in another. }
\label{plot03}
\end{figure}

As the elementary charge is carried by a single proton, it is energetically unfavorable when two protons occupy both 
minima of the same H-bond (in hydrogen-bonded systems such an orientational defect is referred as Bjerrum $D$ defect), 
or if both minima are not occupied (Bjerrum $L$ defect) because of interacting electron clouds. 
This is included in our model by introducing an additional Boltzmann factor. 
In the presented model these defects give rise to transition rates only when they appear together (see Fig.~\ref{plot03}),
so without the loss of generality we assume the energies of both defects to be equal to $V_\text{Coul}$
and the corresponding Boltzmann factor is equal to
\begin{equation}
\gamma_C=\exp\left(-\frac{2V_\text{Coul}}{k_B T}\right)\;.
\end{equation}
According to Hassan {\it et al.} \cite{Hassan} the energies of $D$ and $L$ defects for ice are similar and of order 0.4 eV.

For all other situations, including inverse ones to that in Fig.~\ref{plot03}, i.e., those in which before the rotation
two protons occupy both minima in one H-bond and there are no protons in the second H-bond, we put $\gamma_C=1$.
Finally, the transition rate for a rotation, $\gamma_R$, is given by

\begin{equation}
\gamma_R=\nu_R\gamma_C\;, \label{gng}
\end{equation}
where $\nu_R$ is frequency of rotation alone.

\subsection{The relative frequency}

\begin{figure}[htb]
\centering
\includegraphics[width=8.4cm,height=5.19cm]{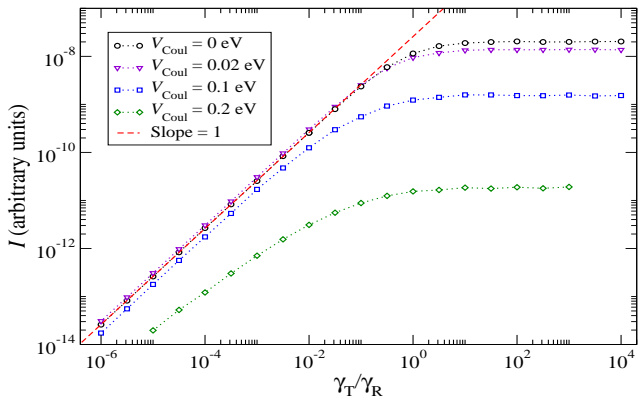}
\caption{(Color online) The log-log dependence of the proton current on the relative frequency
$\gamma_T/\gamma_R$ for the half-filling case $c=0.5$. The individual curves are parametrized by
the Coulomb potential $V_\text{Coul}$.}
\label{plot04}
\end{figure}

The Grotthuss mechanism consists of two kind of processes: the hoppings and the rotations. Thus the behavior
of the current is modeled by the ratio of the characteristic frequencies for hopping ($\gamma_T$) and
rotation ($\gamma_R$). As the the relative frequency varies we observe a nontrivial crossover behavior of
the proton current around $\gamma_T/\gamma_R=1$ (see Fig.~\ref{plot04}). In the rotation-dominated regime
the thick dashed line has slope equal to 1 resulting in the linear dependence of the proton current on the relative
frequency. It is a consequence of the fact that protons are supplied ``on time'' by rotating molecules.
Contrary to this in the tunneling-dominated regime the current saturates within a broad relative frequency
range. This means that when the tunneling frequency is very high, rotating molecules are not able to transfer protons
on quickly enough.

\bigskip
\bigskip

It is worth stressing that although the plot was made for the proton concentration $c=0.5$ a similar 
dependence can be observed for the proton concentration $c\ne 0.5$. The only difference is that far from
$c=0.5$ the dependence on $V_\text{Coul}$ vanishes for $\gamma_T/\gamma_R<1$, while for $\gamma_T/\gamma_R>1$ 
the differences between curves with different values of $V_\text{Coul}$ are reduced by some orders of magnitude 
in comparison to the case with $c=0.5$.

\subsection{Current dependence on the proton concentration}

As one can see in Fig.~\ref{plot04} there is a nonmonotonic dependence of the current with respect to the Coulomb
potential $V_\text{Coul}$ at half-filling. In the tunneling-dominated regime a monotonic decrease of the current with
$V_\text{Coul}$ can be observed whereas the maximal current is for a nonvanishing potential in the
rotation-dominated regime. As one leaves the vicinity of the half filling, then the behavior is monotonic over
a wide range of relative frequency.

\bigskip
\bigskip
\begin{figure}[htb]
\centering
\includegraphics[width=8.5cm,height=5.25cm]{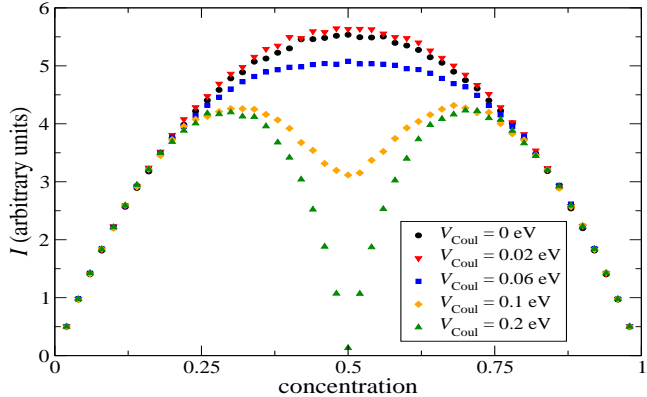}
\caption{(Color online) The current dependence on the proton concentration including the presence
of the Coulomb repulsion for different values of $V_\text{Coul}$. The temperature and the external electric
field are fixed.}
\label{plot05}
\end{figure} 

In order to examine the concentration dependence of the current we fixed the relative frequency at $0.01$ which
naturally means we are in the rotation-dominated regime. As one can see in Fig.~\ref{plot05} the positions of
points are symmetrical about
$c=0.5$, which is a reflection of the particle-hole symmetry in the model.
For $V_\text{Coul}=0$~eV the current has a maximum at $c=0.5$. 
The current slowly rises with the increase of $V_\text{Coul}$ to reach the maximum at about $0.02$~eV which
is of the order of the thermal energy ($T=353$~K in Fig.~\ref{plot05}).
Above this value the local minimum appears at $c=0.5$ instead of maximum
together with two local maxima traveling from $c=0.5$ to approximately $c=0.5\pm0.175$. 
For $V_\text{Coul}>0.1$~eV the minimum goes to zero, while maxima are stable 
in their values. 

This peculiar behavior stems from the fact that the flow of protons is possible when, after 
a rod rotation the proton meets a vacancy on the neighboring rod. This happens when the 
symmetry of the proton arrangement in the chain is not too high. When $V_\text{Coul}=0$ 
and an external electric field is weak, protons (vacancies) have a tendency to form uniform 
clusters, which inhibits proton diffusion. A large value of $V_\text{Coul}$ results in the 
high-symmetry configurations (one proton per rod on the same end of each rod) so the 
presence of protons in the neighboring minima is very unfavorable, which implies a loss 
of current flow. Therefore, a small value of $V_\text{Coul}$ is optimal for a fast 
diffusion. 

This behavior is in agreement with the theoretical predictions derived in the one-dimensional
lattice gas model \cite{Yona} for small values of $V_\text{Coul}$. The initial growth of the current with the proton
concentration is also in agreement with data observed experimentally, e.g. for Nafion, for different
values of hydration \cite{Kreuer2004}. Furthermore, the conductivity for mobile ions in a two-dimensional periodic
potential \cite{Mazu} also exhibits the absolute minimum at $c=0.5$, though it has a richer behavior where more minima
and maxima are present.

\section{Details of dynamics simulations}

The main idea behind the kinetic Monte Carlo method is to use transition rates that depend on the energy barrier
between the states. A technical issue is to choose appropriate method to determine the transitions rates. 
When the rate constants of all processes are known, we can perform the KMC simulations in the time domain. It is
worth noting that in our model the presence of the external electric field modifies rod rotations as well as
proton hoppings. 

\subsection{Rotations}

Herein, the internal rotations of rods are treated as the thermally activated process
satisfying the Arrhenius law

\begin{equation}
\nu_R=\nu^0_R\exp\left(-\frac{V_\text{act}}{k_B T}\right)
\max\left[1,\exp\left(-\frac{|e|Kb}{k_B T}\right)\right]
\;.\label{rr}
\end{equation}
This formula together with Eq. (\ref{gng}) gives the transition rates for rotations.

The last factor represents interaction with the external electric field $K$, $e$ is the elementary charge and $b$---the
size of a rod, $\nu^0_R$ is the frequency of rotation, and $V_\text{act}$ the activation energy for rotation in the absence of the external electric field. We assume that
these values do not depend on temperature. 
The quantity $\nu^0_R$ can be determined by the energy difference of the two lowest states of the quantum rigid rotor
governed by the Schr\"{o}dinger equation

\begin{equation}
\left[-\frac{\hbar^2}{2I}\frac{d^2}{d\phi^2} + V_R(\phi)\right] \psi(\phi) = E \psi(\phi), \label{rotor}
\end{equation}
with the potential
\begin{equation}
V_R(\phi)=\frac{V_\text{act}}{2}\left[1+\cos(2\phi)\right] + |e|Kb \cos(\phi - \phi_0). \label{rot-v}
\end{equation}

The first part of $V_R(\phi)$ is a harmonic twofold potential and the second one describes interaction of a proton with the external 
electric field forming the angle $\phi_0$ with the chain direction. The moment of inertia $I$ depends on the masses and geometry of the molecule. 
It is noteworthy that for a vanishing electric field the solutions of Eq.~(\ref{rotor}) can
be expressed by Mathieu functions. 

Let us note that when changing the angle between the chain and the applied field, then changing the two lowest states
of the quantum rotor. Since the individual chains are distributed randomly in a macroscopic sample, we have to take
this into account.

\subsection{Hopping}

The migration of a proton from one rod to another represents the hopping between the minima of the H-bond potential. 
Hopping is defined as the thermally assisted tunneling which is an extension of the purely classical Arrhenius behavior.
We approximate the H-bond potential by the fuzzy Morse potentials originating in rod ends as they 
represent anionic groups between which the H-bonds are created in real materials. In our model the size of the rod is 
kept fixed while the distance between rods may vary somewhat with temperature.
\begin{widetext}
\begin{eqnarray}
V_a(x)&=&\frac{1}{2a}\int_{-a}^a \left[V_\text{Morse}\left(\frac{d}{2}-x+y\right)
+ V_\text{Morse}\left(x-y-\frac{d}{2}\right)\right] dy\;,\label{va}\\
V_\text{Morse}(x) &=& g \left[\exp\left(-\frac{2x}{b}\right) -2 \exp\left(-\frac{x}{b}\right)\right]\;. \label{pot}
\end{eqnarray}
\end{widetext}
$V_a(x)$ is the single or double well potential but we focus only on the second one in this paper.
The parameter $a$ controls the dispersion in the position of the anionic groups forming the H-bond and it represents
the lattice vibrations (the influence of phonons on the potential). The choice of the Morse potential is dictated by
the fact that it can be very well fitted to H-bond potentials \cite{Duan}, but this does not mean that this choice is
decisive for our considerations (i.e. we could use the Lennard-Jones potential and get similar results).

We assume the thermal dependencies of the $a$ and $d$ parameters, see Eqs. (\ref{va}) and (\ref{pot}),
are linear in the temperature range corresponding to that examined in the experiments.
\begin{eqnarray}
a(T)&=&a_0 + a_1\,(T-T_0)\;,\\
d(T)&=&d_0 + d_1\,(T-T_0)\;.
\label{eeqq}
\end{eqnarray}

The parameters $g$ and $b$ of the Morse potential are fitted in such a way as to get the distance between the minima 
of the double well potential $V_a$ equal to $\Delta x$ together with the height of the barrier equal to $h$.

In the presence of the external electric field, $K$, the potential of the H-bond is modified by the term $|e|Kx$, so
we define
\begin{equation}
V(x)=V_a(x)+|e| K x\;.
\end{equation}
If the external electric field is not too strong $V(x)$ is the double potential.

\begin{figure}[hbt]
\centering
\includegraphics[width=8.5cm,height=5.25cm]{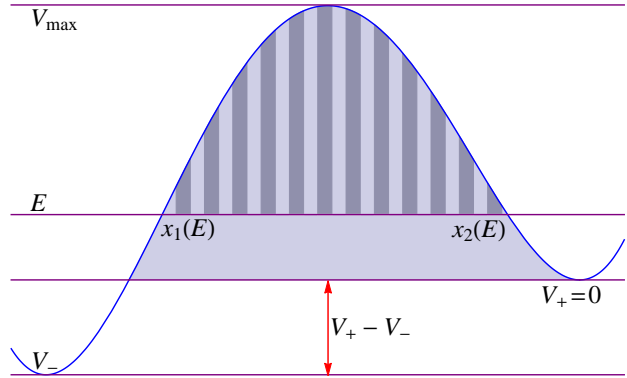}
\caption{(Color online) The shape of the potential $V(x)$ in the presence of an external electric field. 
The dark shaded area (strips) shows the contribution to integral (\ref{wkb}) determined by the value of $E$, 
the brighter one shows the range of energies contributing to Eq. (\ref{tunn}). The hopping from 
the lower minimum, $V_-$, to the upper one, $V_+$, introduces the factor $\exp[-(V_+-V_-)/(k_BT)]$ to overcome 
the physically forbidden region for a proton with energy less than~$V_+$. The energy in Eqs. (\ref{tunn})--(\ref{wkb})
is measured from the upper minimum, i.e., $V_+=0$.}
\label{plot06}
\end{figure}

The tunneling rate is calculated using Bell's formula\footnote{This formula is just the quantum mechanical 
version of the Arrhenius law which is easily seen after rewriting
\begin{displaymath}
\exp\left(-\frac{E_{act}}{k_B T}\right) = 
\frac{1}{k_B T}\int^\infty_0\theta(E-E_{act}) \exp\left(-\frac{E}{k_B T}\right) d E
\end{displaymath}
and replacing the classical Heaviside function $\theta(E-E_{act})$ by the quantum permeability $G(E)$.}\cite{Bell}
\begin{equation}
\tau_T=\frac{1}{k_B T}\int^\infty_0 G(E) \exp\left(-\frac{E}{k_B T}\right) d E \;,\label{tunn}
\end{equation}
with \cite{Kemble}
\begin{equation}
G(E)=\begin{cases}
1/[1+G_\text{WKB}^{-1}(E)]\;, &\text{for}\; E\le V_\text{max}\;,\\
1\;, & \text{for}\; E>V_\text{max}\;,
\end{cases}
\end{equation}
where 
\begin{equation}
G_\text{WKB}(E)=\exp\left(-\frac{2}{\hbar}\int_{x_1(E)}^{x_2(E)}\sqrt{2m[V(x)-E]}\,d x\right) \label{wkb}
\end{equation}
is the WKB quantum permeability of the proton with energy $E$ traveling between classical return points 
$x_1(E)$ and $x_2(E)$ of the potential $V(x)$, see Fig.~\ref{plot06}. Thus, the calculation of the 
tunneling rate $\tau_T$ requires two successive one-dimensional integrations.

When $K\ne 0$ the minima of $V(x)$ have different energies. The proton located at the lower minimum, $V_-$, 
cannot tunnel to the upper one, $V_+$. To take this into account we introduce the extra Boltzmann factor
$\exp[-(V_+-V_-)/k_BT]$  for the hop from the lower to the upper minimum in addition to the tunneling
rate (\ref{tunn}) which represents the tunneling rate for the hop from the upper to the lower minimum.
Thus, the total hopping rate becomes

\begin{equation}
\gamma_T=\nu^0_T\tau_T \times\begin{cases}
1\;, & \text{hopping from $V_+$ to $V_-$}\;,\\
\exp\left(-\frac{V_+-V_-}{k_BT}\right)\;, & \text{hopping from $V_-$ to $V_+$}\;. 
\end{cases} \label{e14}
\end{equation}

The form of Eq. (\ref{e14}) ensures that the detailed balance is fulfilled because it is of the {\it Metropolis}-like rate
type \cite{Fic91}.

\subsection{Finite-size effects}

\begin{figure}[htb]
\centering
\includegraphics[width=8.5cm,height=5.25cm]{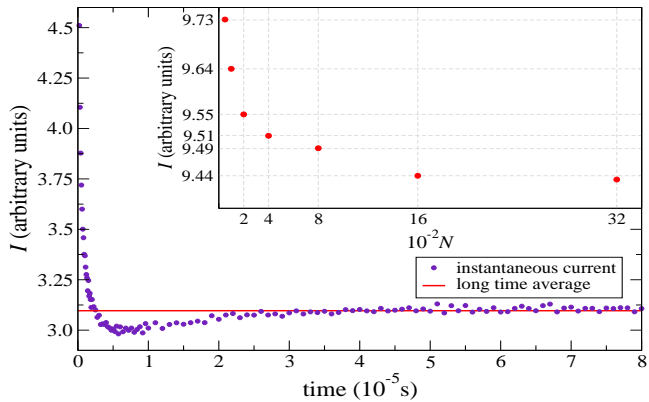}
\caption{(Color online) The current flow stabilizes after less than $5\times 10^{-5}$ s, thus the switch-on
effect may be neglected. Inset: the exemplary results for different system sizes converging quickly with the system
size  (the first two points are for $N=50$ and 100).}
\label{plot07}
\end{figure}

The current was measured by counting the protons hopping from rod $N/2$ to rod $N/2+1$, where $N$ is 
the length of the chain, minus the number of protons moving in the opposite direction during the time of calculations.
The initial configuration was randomly chosen and the final result for the proton current was the average of several
initial configurations. Such a small number of initial configurations was good enough because the saturation time was
much less than the time needed to observe the current flow, Fig.~\ref{plot07}.

\bigskip

The number of KMC steps during an individual program run was of the order $10^7$--$10^9$ (0.01--1 s of the time evolution)
which gave several hundred protons counted to yield the  value of the current with the numerical accuracy better
than 5\%.

In the inset of Fig.~\ref{plot07} the dependence on the chain length, $N$, is presented confirming that finite-size
effects become negligible for larger systems.

\section{Benzimidazole as an example of model implementation}

The benzimidazole belongs to the large family of heterocycles, which are possible alternative material for 
membranes functioning in the intermediate operating temperature range \cite{pub01,pub02,pub03,pub04}. 
The crystal structure of the polycrystalline benzimidazole \cite{pub05,pub06,pub07,pub08} revealed the hydrogen 
bond formation of the N--H$\cdots$N type (with hydrogen bond distances of 2.885~\AA) among the adjacent benzimidazole 
molecules. The H-bond is almost linear (the angle $\measuredangle$(NHN)$=172^\circ$ \cite{pub07}) thus, 
our description by the one-dimensional potential is reasonable.
The characteristic structural features of the benzimidazole crystal are parallel two-dimensional layers. 
In each layer one can distinguish the infinite ribbons made of benzimidazole molecules linked by the N–H$\cdots$N 
hydrogen bridge that play the role of the conducting paths.

According to impedance spectroscopy and 1H NMR experimental results \cite{MZ} the proton conduction process of
the benzimidazole can be considered as a cooperative one involving both molecular motions prior to the 
proton exchange and migration along the hydrogen bonded chain via the N--H$\cdots$N bridges. The first process 
occurs due to the $180^\circ$ flip of a bicyclic molecule (the fusion of benzene and imidazole) which was confirmed 
in experimental studies of the 1H NMR second moment temperature dependence \cite{MZ}. For this reason, it should 
be well described by our model system of rods each of which has only two positions. In addition, the well-known 
structure of the benzimidazole crystal makes it an excellent model molecular system for investigation of the electric 
conductivity process efficiency at the microscopic level. The benzimidazole was chosen as the proton carrying compound 
also due to high chemical and thermal stability. Benzimidazolium cations do not diffuse in the bulk of the sample 
even near melting temperature.

We are going to test our model by comparing experimental results and computer simulations for the electrical conductivity
of the benzimidazole, where the proton concentration is $1/2$. The moment of inertia of the benzimidazole molecule is 
calculated with respect to the longitudinal axis around which the molecule flips through 
$\pi$ radians. Moreover the rods length, $b$ can be accurately determined by the geometry of the benzimidazole molecule. The values of all parameters used for simulations are given in Table \ref{tab1}.
The system size for simulations $N = 400$ is large enough to avoid finite-size effects.

\begin{figure}[htb]
\centering
\includegraphics[width=8.5cm,height=5.25cm]{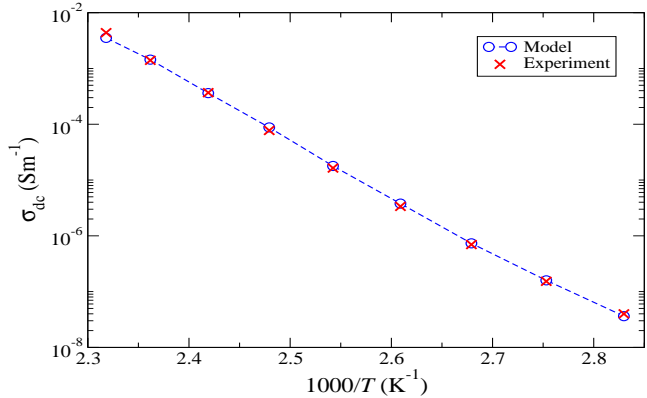}
\caption{(Color online) Comparison between the measured and simulated data for the benzimidazole.}
\label{plot08}
\end{figure}

The electric conductivity measurements of the benzimidazole were carried out by means of impedance spectroscopy using
a Novocontrol Alpha A Frequency Analyzer in the frequency range from 1~Hz to 10~MHz. The real resistance of
the material was evaluated by a fitting procedure using the parallel $RC$ equivalent circuit model.
The current (the $\sigma_{dc}$ conductivity)
of the sample calculated from its bulk resistance $R$ is displayed as a function of inverse temperature in
Fig.~\ref{plot08} (crosses). Measurements were made in the temperature range, from 353~K to above 431~K, near
the melting point. The temperature of the sample was stabilized to the accuracy of 0.01~K using a Novocontrol
Quatro Cryosystem.

\bigskip
\bigskip

\begin{table}[thb]
\caption{Values of parameters for benzimidazole simulations.}
\begin{tabular}{@{}*{4}{l}}
\hline
\hline
Parameter  & Symbol  & Value & Derivation \\
\hline
Frequency of rotation prefactor\footnote{For the  benzimidazole ($V_\text{act}=0.269$ eV)
the lowest states are degenerated forming doublets when the electric field 
$K$ is zero. The value of $\nu_R^0$ is determined by the energy difference between the lowest two doublets.
When the electric field is non-zero then the degeneracy is intact for $\phi_0=0,\pi$ and $\nu_R^0$ 
changes only slightly. When $\phi_0\ne 0,\pi$ the degeneracy 
is quickly removed and $\nu_R^0$, calculated now from the energy difference of two lowest states,
reaches the maxima for $\phi_0=-\pi/2,\pi/2$. 
Fortunately, it turns out that $\nu_R^0$ for the electric field perpendicular to the chain of rods ($\phi_0=-\pi/2,\pi/2$) 
is almost equal to the parallel case ($\phi_0=0,\pi$). Therefore, for simplicity, we assume that the electric field 
is always parallel to the chain axis.}
& $\nu_R^0$        & $10^{12}$ Hz  & Eq. (\ref{rotor})\\
Activation energy for rotations & $V_\text{act}$  & 0.269 eV & Ref. \cite{MZ}\\
Rods length                       & $b$              & 3.84 \AA & Ref. \cite{pub07}\\
Moment of inertia                 & $I$              & 123.6 u \AA$^2$ & Ref. \cite{pub07}\\
External electric field\footnote{There is a linear response regime.}
& $K$ & 0.005 $V/$\AA  & \\
Bond length & $d_0$            & 2.886 \AA & Ref. \cite{pub07}\\
Thermal expansion coefficient & $d_1$            & 1.1$\times 10^{-5}$ \AA$/$K & Ref. \cite{PME}\\
$V_a$ barrier height & $h(T_0)$         & 0.38 eV & Ref. \cite{Duan}.\\
Distance between minima of $V_a$ & $\Delta x(T_0)$  & 0.77 \AA & Ref. \cite{Duan}.\\
Reference temperature             & $T_0$            & 393 K \\
$D$ and $L$ defects energy & $V_\text{Coul}$ & 0.04 eV & Fitted, see Fig.~\ref{plot08}\\
Frequency of hopping prefactor   & $\nu_T^0$        & $10^{9}$ Hz & Fitted, see Fig.~\ref{plot08}\\
Lattice vibration amplitude        & $a_0$            & 0.2 \AA & Fitted, see Fig.~\ref{plot08}\\
Thermal susceptibility of $a$    & $a_1$            & 0.002 \AA$/$K & Fitted, see Fig.~\ref{plot08}\\
\hline
\hline
\end{tabular}
\label{tab1}
\end{table}

What is characteristic of the benzimidazole is that its conductivity increases rapidly as is the case 
in our measurements, wherein the current increases by five orders of magnitude in the temperature range of 80 K. 
As can be inferred from
Fig.~\ref{plot04}, such a huge increase in conductivity must be due to a significant change of the relative frequency
$\gamma_T/\gamma_R$. The rotation frequency, for this fairly narrow temperature range, varies no more than an
order of magnitude. Thus, the change in the relative frequency can only be the result of changes in the tunneling
frequency. As the barrier height of the H-bond potential grows with the bond distance [the parameter $d$
in Eq.~(\ref{eeqq})], the only way to lower this barrier and increase the tunneling frequency is to account the thermal
lattice vibrations. Due to vibrations the Morse potential barrier is lowered effectively and a current flows more easily. 
Indeed, the value of $a$ of the order 0.2--0.3 \AA{} can cause changes in $\nu_T$ even as six orders of magnitude. 
Therefore, the role of parameter $a$, responsible for the thermal lattice vibrations, proved to be crucial.

The frequency of tunneling depends on $\nu_T^0$ and the shape of the potential $V_a$ determined by the 
six parameters: $g$, $b$, $d_0$, $d_1$, $a_0$ and $a_1$ [see Eqs.~(\ref{va}),(\ref{pot})].
The parameters $d_0$ and $d_1$ are known while $g$ and $b$ can
be fitted directly from the Duan analysis \cite{Duan,Schuster} carried out for the parametrization of N--H$\cdots$N potential
at the temperature $T=393$ K. Thus, only three parameters responsible for the frequency of tunneling $\nu_T^0$, $a_0$, $a_1$, and
$V_\text{Coul}$, related to the dynamical modification of the frequency of rotations, are free.
Fortunately, we were able to set physically meaningful values of these parameters
to get a very good agreement with experimental data (see Table \ref{tab1} and Fig.~\ref{plot08}). The mutual interplay between $d_1$,
$a_0$, $a_1$ and the ratio $\gamma_T/\gamma_R$ is responsible for the concavity of the simulation curve. 

\section{Conclusions}

The proton conduction is of outstanding importance for a wide range of technologically significant processes. 
Its theoretical description provides a challenge since it comprises classical and quantum transport phenomena.
We have proposed a microscopic model of the proton conductivity based on the kinetic Monte Carlo approach
adequate to characteristic time scales for the proton conduction. It has been examined that our one-dimensional
model can describe qualitatively and quantitatively the proton diffusion in anhydrous proton conductors.

Generally the proton conducting polymers can be divided into two types: hydrous proton conducting polymers with
a solvent assisted proton transfer and anhydrous ones where protons are transferred via the Grotthuss mechanism.
The latter, similarly as the solid acids, can operate at high temperature (above the water boiling point) and
are the main object of our interest. We have implemented the two-stage Grotthuss proton migration mechanism into
our model and showed in detail that the relative frequency of reorientation and diffusion processes is crucial
for the proton conductivity.

Our model has been applied successfully to describe the proton transport in the polycrystalline benzimidazole. 
It is worth stressing that most of the parameters have been estimated
on the basis of experimental data or the quantum-mechanical calculations. Our simulations of the proton current have
demonstrated not only the very good agreement with the experimental data, but furthermore, proved that the thermal
lattice vibrations, which modify the H-bond potential, play an essential role in the conduction process.

In our opinion the proposed model could be extended in several directions. First, it could be applied to at least
some of other anhydrous proton conductors including two- or three-dimensional systems. Second, our model can be used
to examine effects of hydrostatic pressure elevation---our preliminary results for the benzimidazolium azelate are promising.
Another attractive perspective is the study of the alternating current conductivity.

\theendnotes

\begin{acknowledgments}
This work was supported by the Polish Ministry of Science and Higher Education through Grant No. N N202 368139.
\end{acknowledgments}

\end{document}